\documentstyle[subfigure,preprint,aps,epsfig]{revtex}
\begin{document}
\preprint{KAIST-TH 2001/09}
\title{SUSY CP violations in $K$ and $B$ systems}

\author{P. Ko \footnote{To appear in the proceeding of CICHEP 2001}}

\address{Dep. of Physics, KAIST, Taejon 305-701, Korea 
\\E-mail: pko@muon.kaist.ac.kr
}


\maketitle
\begin{abstract}
I report our recent works on the effects of 
CP violating phases in SUSY models on $B$ and $K$ phenomenology.  
\end{abstract}

\tighten
\vspace{.2in}

\section{Introduction}
In the minimal supersymmetric standard model (MSSM), there arise many new
CP violating (CPV) phases beyond the KM phase in the standard model (SM). 
These SUSY CPV phases generically lead to too large electron/neutron 
electric dipole moment (EDM) and $\epsilon_K$. There are basically three
different ways to evade this SUSY CP problem :\cite{susycp} 
(i) vanishing (or very small) SUSY CP phases, 
(ii) decoupling scenario (effective or more minimal MSSM $\equiv$ MMSSM), 
and (iii) cancellation mechanism.\cite{nath}  
In each case, these new SUSY CP phases could affect $B$ and $K$ physics 
in vastly different manners. 

In this talk, we report our three recent works related with this  
subject.\cite{ko1,ko2,ko3}  The topics covered here are divided into 
two parts. 
In the first part, we consider a possibility of observing effects 
of the $\mu$ and $A_t$ phases which are flavor conserving  
at $B$ factories in the MMSSM (including the electroweak baryogenesis
scenario therein \cite{carena}).
In the second part of the talk, I demonstrate that fully supersymmetric CP 
violations in the kaon system are possible, on the contrary to the
conventional folklore. Both $\epsilon_K$ and 
${\rm Re}(\epsilon^{'} / \epsilon_K)$ 
can be accounted for in terms of a {\it single} CP violating complex number 
$( \delta_{12}^d )_{LL}$ (to be defined below) within the double  mass 
insertion  approximation (MIA).

\section{Effects of $\mu$ and $A_t$ phases on $B$ physics in the more 
minimal supersymmetric standard model (MMSSM)}

In the MMSSM (decoupling sceario), 
the 1st and the 2nd family squarks are assumed to be very heavy and almost 
degenerate in order to solve the SUSY FCNC/CP problems.\cite{kaplan}  
Only the third family squarks can be light enough to affect
$B\rightarrow X_s \gamma$ and $B^0 - \overline{B^0}$ mixing.
We also ignore possible flavor changing squark mass matrix elements
that could generate gluino-mediated flavor changing neutral current (FCNC)
process in addition to those effects we consider below. Recently, such 
effects were studied in the $B^0 - \overline{B^0}$ 
mixing,\cite{cohen-B,randall} the branching ratio of 
$B\rightarrow X_s \gamma$ \cite{cohen-B} and CP violations 
therein,\cite{hou,kkl} and $B\rightarrow X_s l^+ l^-$,\cite{kkl} respectively. 
Ignoring such contributions, the only source of the FCNC in our model is the 
CKM matrix, whereas there are new CPV phases coming from the phases of $\mu$ 
and $A_t$ parameters in the flavor preserving sector in addition to the KM 
phase $\delta_{KM}$ in the flavor changing sector. (We choose to work in the 
basis where the wino mass parameter $M_2$ is real.) In this sense, this paper
is complementary to the ealier works.\cite{cohen-B,randall,hou,kkl}

Even if the 1st/2nd generation squarks are very heavy and degenerate, there
is another important edm constraints considered by Chang, Keung and Pilaftsis 
(CKP)  for large $\tan\beta$.\cite{pilaftsis}  
This constraint comes from the two loop diagrams involving stop/sbottom 
loops, and is independent of the masses of the 1st/2nd generation squarks.
Therefore, this CKP edm constraints can not be simply evaded by making the 
1st/2nd generation squarks very heavy, and it turns out that this puts a 
very strong constraint on the possible new phase shift in the 
$B^0 - \overline{B^0}$ mixing.  

\begin{figure}
\centerline{\epsfxsize=10.3cm \epsfbox{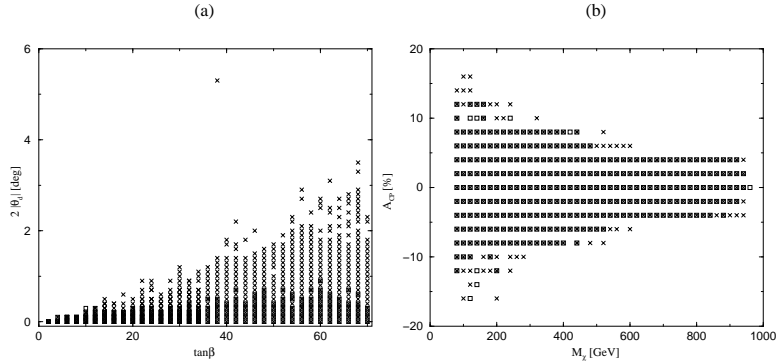}}
\caption{
Correlations between
(a) $\tan\beta$ vs. 2 $|\theta_d | $, and 
(b) ${\rm Br} (B \rightarrow X_s \gamma)$ vs. 
$A_{12}^{\rm FULL} / A_{12}^{\rm SM}$. 
The squares (the crosses) denote those which (do not) satisfy the CKP edm 
constraints.  
}
\label{fig1}
\end{figure}

The $B^0 - \overline{B^0}$ mixing is generated by the box diagrams 
with $u_i-W^{\pm} (H^{\pm})$ and $\tilde{u}_i-\chi^{\pm}$ running around 
the loops in addition to the SM contribution.\cite{branco} 
The gluino and neutralino contributions are negligible in our model.
It turns out that the chargino exchange contributions to 
$B^0 - \overline{B^0}$ mixing can be generically complex relative to the 
SM contributions, and  can generate a new phase shift in the 
$B^0 - \overline{B^0}$ mixing relative to the SM value. This effect can be 
in fact significant for large $\tan\beta (\simeq 1/\cos\beta)$, since the 
chargino contribution  is proportional to $ (m_{b} / M_W \cos\beta )^2$. 
\cite{demir}
However, the CKP edm constraint puts a strong constraint for large 
$\tan\beta$ case. 
In Fig.~\ref{fig1} (a),
we plot $ 2 \theta_d \equiv {\rm Arg}~(M_{12}^{\rm FULL} / 
M_{12}^{\rm SM} )$ as a function of $\tan\beta$.
The open squares (the crosses) denote those which (do not) satisfy the
CKP edm constraints.
It is clear that the CKP edm constraint on $2 \theta_d$ is in fact very
important  for large $\tan\beta$, and we have $| 2 \theta_d | \leq 1^{\circ}$.
This observation is important for the CKM phenomenology,
since time-dependent CP asymmetries in neutral $B$ decays into
$J/\psi K_S, \pi\pi$ etc. would still measure directly three angles of the
unitarity triangle even in the presence of new CP violating phases,
$\phi_{A_t}$ and $\phi_{\mu}$.
The above $B^0 - \overline{B^0}$  mixing is also related with the dilepton
asymmetry which is proportional to ${\rm Re} ( \epsilon_B )$.
Neglecting the small SM contribution which is about $\sim 10^{-3}$,
we found that $| A_{ll} | \leq 0.1 \%$ in our mode, which  is well
below the current data, $A_{ll} = (0.8 \pm 2.8 \pm 1.2 ) \%$. \cite{opal}
On the other hand, if any appreciable amount of the dilepton asymmetry is
observed, it would indicate some new CPV phases in the off-diagonal
down-squark mass matrix elements,\cite{randall}
assuming the MSSM is realized in nature.

Unlike the $\theta_d$ and $A_{ll}$ discussed in the previous 
paragraphs, the magnitude of $M_{12}$ can be affected a lot by the $\mu$ and
$A_t$ phases, 
and it will affect the determination of $V_{td}$ from $\Delta m_{B_d}$. 
The deviation from the SM can be as large as $\sim 60 \%$,\cite{ko2} and the 
correlation behaves differently from the minimal supergravity case.\cite{goto}
  
The radiative decay of $B$ mesons, $B\rightarrow X_s \gamma$, is
described by the effective Hamiltonian including (chromo)magnetic dipole
operators. Interference between $b\rightarrow s \gamma$ and $b\rightarrow 
s g$ (where the strong phase is generated by the charm loop via 
$b\rightarrow c\bar{c}s$ vertex) can induce direct CP violation in 
$B\rightarrow X_s \gamma$.\cite{KN} 
In our model, $A_{\rm CP}^{b\rightarrow s \gamma}$ can be as large 
as $\simeq \pm 16\%$ if chargino is light enough, even if we impose all 
the relevant edm  constraints \cite{ko1} (see Fig.~1 (b)). 

The $\mu$ and $A_t$ phases can also affect $B\rightarrow X_s l^+ l^-$, 
whose branching ratio can be enhanced upto 
$\sim 85 \%$ compared to the SM prediction in our model,\cite{ko2}   
and the correlation between ${\rm Br} (B\rightarrow X_s \gamma)$ and 
${\rm Br} (B\rightarrow X_s l^+ l^-)$ is distinctly different from the 
minimal supergravity scenario (CMSSM) (even with new CP violating phases)
\cite{okada} in the presence of new CP violating phases in $C_{7,8,9}$
as demonstared in model-independent analysis by Kim, Ko and Lee.\cite{kkl}
It is crucial to consider the direct CP asymmetry in $B\rightarrow X_s \gamma$
in the model independent determination of $C_{7,9,10}$, since these Wilson
coefficients may be complex in general. 

One can also consider the $K^0 - \overline{K^0}$ mixing in this model, and  
we find $\epsilon_K / \epsilon_{K}^{SM}$ can be as large as 1.4 for
$\delta_{KM} = 90^{\circ} $.\cite{ko2}
 This is a factor 2 larger deviation from the 
SM compared to the minimal supergravity case.\cite{goto} 
The dependence on the lighter stop is close to the case of the 
minimal supergravity case, but we can have a larger deviations. 
Such deviation is reasonably close to the experimental value, and will 
affect the CKM phenomenology at a certain level. 
This is the extent to which the new 
phases in $\mu$ and $A_t$ can affect the construction of the unitarity 
triangle through $\epsilon_K$. 

\section{Fully SUSY CP violation in the kaon system}

In the MSSM with many new CPV phases, there is an intriguing possibility that 
the observed CP violation in $K_L \rightarrow \pi\pi$ is fully supersymmetric 
due to the complex parameters $\mu$ and $A_t$ in the soft SUSY breaking terms 
which also break CP softly, or CP violating $\tilde{g} - q_i - \tilde{q}_j$. 
Our study indicates that the supersymmetric $\epsilon_K$ (namely, for 
$\delta_{KM} = 0$) is less than $\sim 2\times 10^{-5}$, which is too small 
compared to the observed value : $ | \epsilon_K | = (2.280 \pm 0.019) \times 
10^{-3}$.
Details for this discussions can be found in Ref.~\cite{ko2}.

Although one cannot generate enough CP violations in the kaon system through
flavor preserving $\mu$ and $A_t$ phases in the MSSM, it is possible if one
considers the flavor changing SUSY CPV phases. Let us work in the mass 
insertion approximation (MIA) and consider the flavor changing 
$\tilde{g}-q_{i}-\tilde{q_j}$ in order to study these pheomena in a model 
independent manner. The folklore was that if one saturates the $\epsilon_K$
with $(\delta_{12}^d )_{LL}$, the corresponding $\epsilon^{'} / \epsilon_K$ 
is far less than the observed value. On the other hand, if one saturates 
$\epsilon^{'} / \epsilon_K$ with $(\delta_{12}^d )_{LR}$, the resulting 
$\epsilon_K$ is again too small compared to the data. Therefore one would 
need both 
$| (\delta_{12}^d )_{LL} | \sim O( 10^{-3} ) ~~~~~{\rm and}~~~~~
| (\delta_{12}^d )_{LR} | \sim O( 10^{-5} )$, 
each of which has a $\sim O(1)$ phase. Masiero and Murayama argued 
that  such a large value of $ (\delta_{12}^d )_{LR}$ is not implausible in
general MSSM, e.g., if the fundamental theory is a string theory.\cite{mm}  
In their model, the large $(\delta_{12}^d )_{LR}$ is intimately related 
with the large $(\delta_{11}^d )_{LR}$, so that their prediction on the 
neutron edm is very close to the current experimental limit. 

Subsequently, we pointed out that a single complex number
$(\delta_{12}^d )_{LL} \sim  O ( 10^{-2} - 10^{-3} )$ with an order 
$\sim O(1)$ phase in fact can generate both 
$\epsilon_K$ and $\epsilon^{'} / \epsilon_K$,  
if one goes beyond the single mass insertion approximation
as often done in this field.\cite{ko3} In our model, 
$\epsilon_K$ is generated by 
$(\delta_{12}^d )_{LL}$, whereas $\epsilon^{'} / \epsilon_K$ is generated 
by a flavor preserving $\tilde{s}_R - \tilde{s}_L$ transition followed by 
flavor changing $\tilde{s}_L - \tilde{d}_L$ transition. 
\begin{figure}
  \begin{center}
      \begin{picture}(500,130)
        \centerline{\epsfxsize=6cm \epsfbox{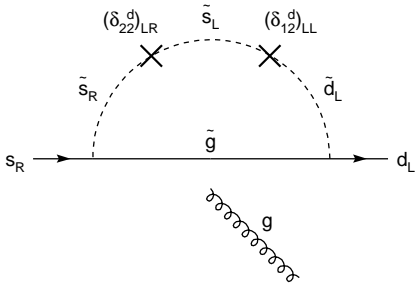}}
        \end{picture}
    \caption{Feynman diagram for \(\Delta S = 1\) process.}
    \label{fig:sdglue}
  \end{center}
\end{figure}
The former is proportinal to $m_s ( A_s - \mu \tan \beta ) / \tilde{m}^2$
where $\tilde{m}$ is the common squark mass in the MIA. This kind of the 
induced $LR$ mixing is always present generically in any SUSY models.

If the KM phase were not zero,
the CKMology should differ significantly from the SM case. For example,
we cannot
use the constraints coming from $\epsilon_K$ or $\Delta M_{B}$, since new
physics would contribute to both $\Delta S= 2$ and $\Delta B= 2$ amplitudes.
More detailed discussions on these points will be
presented elsewhere.\cite{ko4}
Finally let us note that the recent
observation on CP asymmetry in $B^0 \rightarrow J/\psi K_S$ depends on
different CP violating parameter $( \delta_{i3}^d )_{AB}$ where $i=1$ or $2$,
and $A,B = L$ or $R$, and is independent of the kaon sector we considered
here in the mass insertion approximation.

\begin{figure}
  \begin{center}
    \begin{picture}(700,230)
      \centerline{\epsfxsize=8.3cm \epsfbox{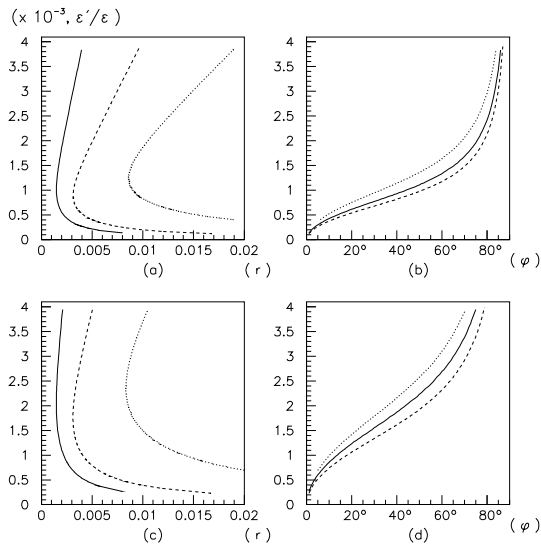}}
    \end{picture}
    \caption{$\epsilon' / \epsilon_K$ as a function of the modulus $r$
      [(a) and (c)] and the phase $\varphi$ [(b) and (d)] of the parameter 
      $(\delta_{LL}^d )_{12}$ 
      with $A_s - \mu^* \tan \beta $ to be $-10~TeV$ [(a),(b)] 
      and $-20~TeV$ [(c),(d)].  
      The common squark mass is chosen to be $\widetilde{m} = 500$ GeV, 
      and the solid, 
      the  dashed and the dotted curves correspond to $x = 0.3, ~1.0, ~2.0$, 
      respectively.}
    \label{fig:epsp}
  \end{center}
\end{figure}

\section{Conclusion}
In conclusion, we first discussed the effects of $A_t$ and $\mu$ phases 
in the MMSSM 
on several observables in the $B$ meson system and on
$\epsilon_K$.  Our study includes the EW baryogenesis scenario in the MSSM.
We demonstrated that the SUSY CP violating phases in $\mu$ and $A_t$ can 
affect the $B$ and $K$ physics a lot (especially direct CP asymmetry in 
$B\rightarrow X_s \gamma$), even if they are flavor conserving. 
We also argued that if we consider SUSY phases with flavor violation such 
as a flavor changing squark 
mixing $( \delta_{12}^d )_{LL}$, fully supersymmetric CP violation is 
possible for relatively large $\tan\beta$ (more specifically, 
$| \mu \tan \beta | \sim O(10)$ TeV ) in the double mass insertion 
approximation. 
The ongoing $B$ and $K$ factory experiements will test the KM paradigm for 
CP violation, and will shed light on the possible new 
sources of CP violations (which is necessary in the electroweak baryogenesis 
scenario), 
especially if the MSSM is the correct effective theory for nature at the 
electroweak scale.

\section*{Acknowledgments}
The author is grateful to S. Baek, J.-H. Jang, Y.G. Kim, J.S. Lee and 
J.H. Park for collaborations on the works presented in this 
talk, and also to Shaaban Khalil for the nice organization of 
this conference.   
This work is supported in part by BK21 program of Ministry of Education and 
by SRC program by KOSEF through CHEP at Kyungpook National University.


\begin{thebibliography}{99}
\bibitem{susycp} See, for example, Y. Nir, 
Lectures given at 27th SLAC Summer Institute on Particle Physics: CP
Violation in and Beyond the Standard Model (SSI 99),
Stanford, California, 7-16 Jul 1999. hep-ph/9911321.
\bibitem{nath} T. Ibrahim and P. Nath, Phys. Lett. {\bf B 418}, 98 (1998) ;
Phys. Rev. {\bf D 57}, 478 (1998) ; (E) {\it ibid.}, {\bf D 58}, 019901 
(1998) ; 
Phys. Rev. {\bf D 58}, 111301 (1998) ;
M. Brhlik, G.J. Good and G.L. Kane, Phys. Rev. {\bf D 59}, 115004 (1999).
\bibitem{ko1} S. Baek and P. Ko, Phys. Rev. Lett. {\bf 83}, 488 (1999).
\bibitem{ko2} S. Baek and P. Ko, 
Phys. Lett {\bf B 462}, 95 (1999). 
\bibitem{ko3} S. Baek, J.-H. Jang, P. Ko and J.H. Park, 
Phys. Rev. {\bf D 62}, 117701 (2000). 
\bibitem{carena} 
M. Carena, M. Quiros and C.E.M. Wagner, Nucl. Phys. {\bf B 524}, 3 (1998) ;
J.M. Cline and G.D. Moore, Phys. Rev. Lett. {\bf 81}, 3315 (1998).
\bibitem{kaplan} A.G. Cohen, D.B. Kaplan, A.E. Nelson, Phys. Lett. 
{\bf B388}, 588 (1996).
\bibitem{cohen-B}  A. G. Cohen, David B. Kaplan, F. Lepeintre, 
Ann E. Nelson, Phys. Rev. Lett.{\bf 78}, 2300 (1997).  
\bibitem{randall} L. Randall and S. Su, Nucl.Phys. {\bf B 540}, 37 (1999) ;
G. Barenboim and M. Raidal, Phys.Lett. {\bf B 457}, 109 (1999).
\bibitem{hou} C.-K. Chua, X.-G. He and W.-S. Hou, hep-ph/9808431.
\bibitem{kkl} Y.G. Kim, P. Ko and J.S. Lee, 
Nucl. Phys. {\bf B 544}, 64 (1999).
\bibitem{pilaftsis} D. Chang, W.-Y. Keung and A. Pilaftsis,
Phys. Rev. Lett. {\bf 82}, 900 (1999).
\bibitem{branco} G.C. Branco, G.C. Cho, Y. Kizukuri and N. Oshimo, 
Phys. Lett. {\bf B 337}, 316 (1994) ; Nucl. Phys. {\bf B 449}, 483 (1995).
\bibitem{demir} D.A. Demir, A. Masiero and O. Vives, 
Phys. Rev. Lett. {\bf 82}, 2447 (1999). 
\bibitem{opal} K. Ackerstaff {\it et al.}, OPAL Collaboration, Z. Phys. 
{\bf C 76}, 401 (1997).
\bibitem{goto} T. Goto, T. Nihei and Y. Okada, Phys. Rev. {\bf D 53},
5233 (1996).
\bibitem{KN} A. Kagan and M. Neubert, Phys. Rev. {\bf D 58}, 094012 (1998) ;
Eur. Phys. J. {\bf C 7}, 5 (1999).
\bibitem{okada} T. Goto, Y. Okada and Y. Shimizu,
Phys. Rev. {\bf D 58}, 094006 (1998).
\bibitem{mm} A. Masiero and H. Murayama,  Phys. Rev. Lett. 
{\bf 83}, 907 (1999).
\bibitem{ko4} S. Baek, J.-H. Jang, P. Ko and J.H. Park,
KAIST-TH 2001/06.
\end{thebibliography}
\end{document}